\documentclass[aps,pra,twocolumn,groupedaddress,showpacs,amsfonts,amssymb]{revtex4}

\usepackage{graphics}
\usepackage{epsfig}

\begin{document}


\title{Entanglement required in achieving entanglement-assisted channel capacities}


\author{Garry Bowen}
\email{g.bowen@qubit.org}
\affiliation{Centre for Quantum Computation, Clarendon Laboratory, University of Oxford, Oxford OX1 3PU, United Kingdom.}

\date{\today}

\begin{abstract}
Entanglement shared between the two ends of a quantum communication channel has been shown to be a useful resource in increasing both the quantum and classical capacities for these channels.  The entanglement-assisted capacities were derived assuming an unlimited amount of shared entanglement per channel use.  In this paper, bounds are derived on the minimum amount of entanglement required per use of a channel, in order to asymptotically achieve the capacity.  This is achieved by introducing a class of entanglement-assisted quantum codes.  Codes for classes of qubit channels are shown to achieve the quantum entanglement-assisted channel capacity when an amount of shared entanglement per channel given by, $\mathcal{E}^{\mathrm{Random}}_Q \geq 1 - Q_E$, is provided.  It is also shown that for very noisy channels, as the capacities become small, the amount of required entanglement converges for the classical and quantum capacities.
\end{abstract}


\maketitle

\section{Introduction}

Quantum information theory is a generalization of the classical theory of information transmission and processing, where the encoding of information into a quantum system is taken into account \cite{nielsen}.  The quantum phenomenon of entanglement, when utilized in quantum information theory, allows for uniquely quantum phenomena, such as quantum dense coding \cite{bennett92} and quantum teleportation \cite{bennett93}.  Dense coding was the first demonstration that entanglement could increase the classical communication capacity of a noiseless quantum channel by encoding twice as much information than would be possible without shared entanglement.  This protocol required the use one maximally entangled bipartite system, shared by sender and receiver, per use of the noiseless channel.

If the two ends of a quantum channel share unlimited prior entanglement, then the quantum capacities of the channel are known exactly \cite{bennett99,bennett01a}.  The entanglement assisted classical capacity of a channel, $\Lambda$, is given by,
\begin{equation}
C_E = \max_{\rho} \bigg[ S(\rho) + S(\Lambda \rho) - S\big( (\mathbb{I}\otimes \Lambda) |\phi \rangle \langle \phi|\big) \bigg] \; ,
\label{eqn:C_E}
\end{equation}
where $|\phi\rangle$ is any purification of $\rho$, and $S(\rho)$ the von Neumann entropy of the state $\rho$, given by, $S(\rho) = - \mathrm{Tr} \rho \log \rho$.  Traditionally the logarithm is taken to be base 2, giving the information in \textit{bits}.  The equation is the analogue of Shannon's equation for the classical information capacity of a noisy classical channel \cite{cover}.  The term on the right hand side of Eq.(\ref{eqn:C_E}) has previously been labelled as the von Neumann capacity of a quantum channel, and properties such as additivity have been shown to hold \cite{adami97}.  It is known that if the maximum is obtained for $\rho = \frac{1}{d}\mathbb{I}$, for a $d$-dimensional channel, then dense coding suffices to obtain the given capacity.

The entanglement assisted quantum information capacity for a channel is related to the classical quantity in Eq.(\ref{eqn:C_E}), by \cite{bennett99},
\begin{equation}
Q_E = \frac{1}{2}C_E \; ,
\label{eqn:Q_E}
\end{equation}
and the capacity is given in \textit{qubits}.  The equality is derived by utilizing teleportation and dense coding to give a lower bound to $C_E$ of $2Q_E$, and then bound $Q_E$ from below by $\frac{1}{2}C_E$.

In this paper, we examine whether the capacities in Eqs.(\ref{eqn:C_E}) and (\ref{eqn:Q_E}) can be achieved if the shared entanglement per channel is restricted to a predetermined amount, $0 < \mathcal{E} < \infty$, per use of the channel.  We assume that the amount of entanglement is determined by sharing $m$ copies of a maximally entangled state, per $n$ copies of the channel, with $m/n \rightarrow \mathcal{E}$ as $n\rightarrow \infty$.  Shared pure entangled states that are non-maximally entangled can be converted to maximally entangled states with an equivalent amount of entanglement in the asymptotic limit with vanishing amounts of classical communication \cite{lo99}, and can therefore be considered equivalent.  The question of whether the distillable entanglement of shared mixed entangled resources is interconvertable with vanishing classical communication in the asymptotic limit is yet to be determined.  The two quantities of interest are therefore the minimal amounts of shared entanglement required per channel in order to achieve the entanglement assisted channel capacities in the asymptotic limit.  These quantities are denoted by $\mathcal{E}_C$ and $\mathcal{E}_Q$, for the classical and quantum requirements, respectively, and are defined as the limit, $\lim_{n\rightarrow \infty} \inf \{ \mathcal{E}: R_{\mathcal{E}} = R_{\infty} \}$, where $R_{\mathcal{E}}$ is the channel capacity attainable with an amount of shared entanglement per channel equal to $\mathcal{E}$.

An upper bound on the entanglement required is obtained by introducing a class of entanglement-assisted quantum codes.  These codes are based on the stabilizer formalism, with the exception that the ancillas used to encode the state are replaced by shared maximally entangled states.
The examination of these entangled quantum codes may also give insight into the behavior of degenerate as well as non-degenerate quantum codes.  The introduction of degeneracy into entanglement assisted codes acts to give a lower bound on the capacity of a channel without entanglement.

\section{Quantum Error-Correcting Codes}

Quantum states may be protected against decoherence, by encoding the state in a larger Hilbert space, thereby creating redundancy in the states that is resistant to noise.  The theory of such quantum error correcting codes (QECCs) is a rapidly growing area of research \cite{shor95,steane96,ekert96,gottesman96,bennett96,calderbank96a,knill97,cleve97,shor97,zanardi97,lidar99,knill00,zanardi01}.  The standard method of encoding involves introducing many ancilla quantum states in a known preparation.

A quantum code works by embedding a state in a subspace of a larger Hilbert space, that is invariant under the encoding and decoding operations.  A binary quantum code is designated by three parameters $[ n, k, d]$, where $k$ logical qubits are encoded in $n$ physical qubits, and the code has a distance $d$, which means the code can correct, $t = \frac{1}{2}(d-1)$, errors at unknown locations in the code block.  This means when expanded in terms of error operators $E$, a $t$-error correcting code reverses all errors $E$ that have weight $t$ or less.

For codewords $|i\rangle , |j\rangle$ and errors $E_a,E_b$, it is necessary that,
\begin{equation}
\langle i|E_a^{\dag} E_b|j\rangle = 0 \; ,
\end{equation}
for $i\neq j$, otherwise the error $E_a$ on $|i\rangle$ is indistinguishable from the error $E_b$ on $|j\rangle$, and we would not know which error to correct for.  In the case of \textit{non-degenerate} quantum codes, a sufficient condition is also,
\begin{equation}
\langle i|E_a^{\dag} E_b|j\rangle = \delta_{ab}\delta_{ij} \; ,
\end{equation}
so each error takes the code subspace to mutually orthogonal error subspaces $\mathcal{H}_a = E_a \mathcal{H}_{code}$.  For \textit{degenerate} quantum codes the sufficient condition becomes,
\begin{equation}
\langle i|E_a^{\dag} E_b|j\rangle = M_{ba}\delta_{ij} \; ,
\end{equation}
where, $M_{ba} = \langle i |E_b^{\dag}E_a|i\rangle$, is a Hermitian matrix.

A number of bounds exist for codes, including the quantum Hamming bound and quantum Gilbert-Varshamov bound \cite{ekert96}, the no-cloning bound \cite{bruss98,cerf00}, the quantum Singleton bound \cite{knill97}, and the Rains bound \cite{rains96}.  The quantum Hamming bound only applies for non-degenerate quantum codes, whereas the quantum Singleton bound and the Rains shadow enumerator bound both apply to degenerate and non-degenerate codes.  The linear programming bound applies by converting any additive quantum code to a classical code over $GF(4)$ \cite{calderbank96a}.

\subsection{Linear Quantum Codes}

Linear quantum codes (or stabilizer codes) are obtained by utilizing the group structure of the set of errors acting on the Hilbert space in which the state is encoded \cite{gottesman97}.  Using random stabilizer codes it has been shown that there exist codes that achieve rates arbitrarily close to the non-degenerate quantum Hamming bound in the asymptotic limit.

For a group $G$ acting on a Hilbert space $\mathcal{H}$, the stabilizer of an element of the space, $s \in \mathcal{H}$, denoted by $S$, is the set of elements in $G$ for which $s$ is an eigenvector of eigenvalue 1, under the action of $S$.
In the case of stabilizer codes on qubits, the group in question is referred to as the Pauli group, $G_n$, and consists of the $n$-tensor products of the Pauli matrices.  To make the group easier to deal with we assume, $X = \sigma_x$, $Y = -i\sigma_y$, and $Z = \sigma_z$, which gives, $XY = Z$.  For the group to be closed, we must include the element $-1$, the negative of the identity.  However, this acts trivially on the quantum states, as it simply takes, $\alpha|0\rangle + \beta |1\rangle \rightarrow -\alpha|0\rangle - \beta |1\rangle$, which is the same state modulo the phase.  Hence, taking the subgroup $H = \{ \pm 1 \}$, we can actually assume for the most part we are working with the group modulo the signs, $G_n/H$, with the major exception being the determination whether elements \textit{commute} or \textit{anti-commute}.  Since, $(-g)h = h(-g)$, if $g$ and $h$ commute (similarly for anticommutation), we can generally say that $g$ and $h$ commute or anticommute (in $G_n$) whilst considering the group of errors as a subset of $G_n/H$.

If we take our codespace to be a basis of the stabilized Hilbert space $\mathcal{H}_S$, for an Abelian subgroup $S$ of $G_n$, then it is easy to see that these codewords are unaffected by any error contained in $S$.  Hence, we have a degeneracy in the code for the errors in $S$, these errors do nothing.  The set of errors, $E \in G_n$, that commute with $S$, that is, $gE = Eg$, for all $g \in S$, is known as the centralizer of $S$ in $G_n$ and denoted $Z(S)$.  In the case of the error group $G_n$, there exists an equivalence between the centralizer and normalizer, $N(S)$, of the subgroup $S$, that is, $Z(S) = N(S)$, \cite{nielsen}.  Elements of the normalizer not in the stabilizer, $N(S)\backslash S$, give all the errors that result in a logical error on the encoded qubits, which can be seen by the fact that, $g(E|\psi\rangle) = Eg|\psi\rangle = E|\psi\rangle$, for all $g \in S$ and $|\psi\rangle$ in $\mathcal{H}_S$, and hence $E|\psi\rangle$ is in the codespace of $S$.

For all the errors that anti-commute with at least one element of the stabilizer, then,
\begin{equation}
\langle i|F|j\rangle =\langle i|FE|j\rangle = -\langle i|EF|j\rangle = 0 \; ,
\label{eqn:orth-error}
\end{equation}
for $E$ in the stabilizer, and $F \in G_n\backslash N(S)$, and the error $F$ takes the codewords to subspaces orthogonal to the code subspace.
The act of (complete) decoding gives a map, $\kappa:G_n \rightarrow N(S)$, as it takes all the errors which map the code subspace to orthogonal spaces back to the code subspace.  The map from the logical errors to the decoded qubits, $\phi: N(S) \rightarrow G_k$, is then a homomorphism, where $G_k$ is the group of errors on the message qubits.  Hence, $|N(S)| = |G_k|\cdot |S| = 4^k|S|$, and the errors in $N(S)$ are divided into $4^k$ cosets of equal size, with each coset of errors corresponding to one of the $4^k$ logical errors.

The error correction map $\kappa$ is determined by choice of the particular inverse map $h^{-1}$ that takes a non-degenerate error subspace back to the code space.  We can choose the basis in the error subspace, $\{ |\tilde{k}_E\rangle \}$, for a particular error $h$, such that $h|\tilde{k}\rangle = |\tilde{k}_E\rangle$, which gives all the other errors that map the code space to this error space a grouping according to the logical error in this basis.  Suppose $s$ is an element of the stabilizer, then $h = hs$ on the code space, and hence $\kappa(hs)$ acts as the identity on the code subspace.  A similar mapping occurs for all the logical errors on the code subspace.  Hence, under the mapping $\kappa_{h^{-1}}$, if $g \in N(S) \cong g' \in G_k$ then $hg \cong g' \in G_k$.  Since $hN(S)$ is a coset of $N(S)$ in $G_n$, then the map $\phi \circ \kappa : G_n \rightarrow G_k$, divides $G_n$ up into $4^k$ sets of equal size, with all the members of each set corresponding to a different logical error.

\begin{figure}
\epsfig{file=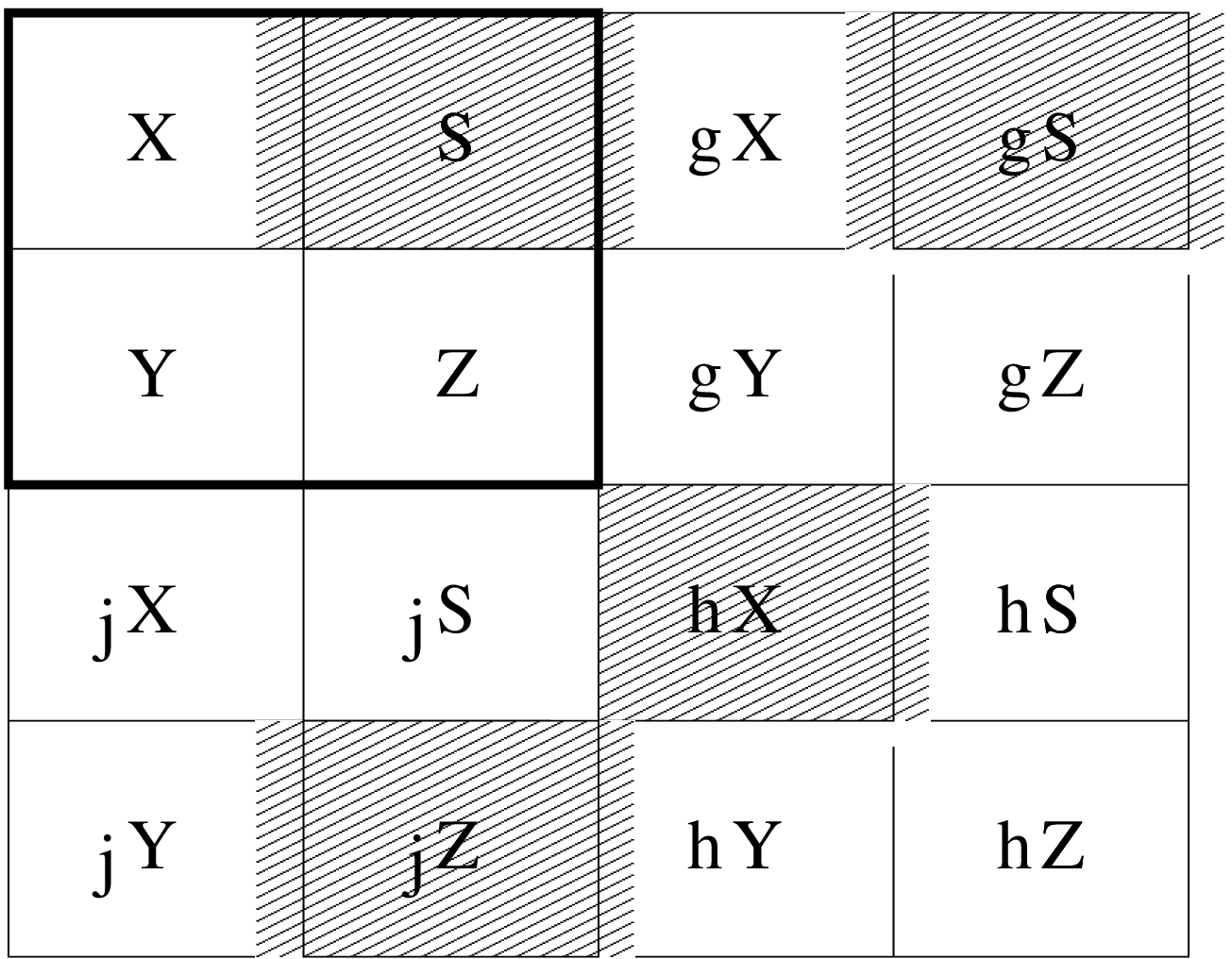,height=6cm}
\caption{Representation of the stabilizer and coset structure for a single encoded qubit.  The stabilizer $S$, and logical errors $X,Y$ and $Z$, form the normalizer of the stabilizer (enclosed in the box).  From each coset of the normalizer $N(S)$, a single coset of the stabilizer $S$ may be mapped back to the stabilizer, representing a correction of the errors in that coset, the rest of the cosets are mapped to the corresponding logical errors.  In this figure the error cosets $gS$, $hX$ and $jZ$ are the corrected cosets of errors.\label{fig:stab01}}
\end{figure}

In summary, a complete error correction scheme determines a map $\kappa$, which we choose to correct a particular member of each coset of $N(S)$, which in turn corrects all members of the image of $S$ in that coset.  Obviously we would like this set to contain the typical errors contained in the given coset of $N(S)$.  A diagrammatic representation of this for a single encoded qubit is shown in FIG. \ref{fig:stab01}.

\section{Entanglement Assisted Codes}

By utilizing part of bipartite entangled states as the ancillas used in coding, the encoder and decoder may be able to create correlations between the encoded state and the reference states held by the receiver.  These correlations enhance the ability of the receiver to decode the state without a logical error on the encoded states, thereby possibly increasing the quantum and classical capacities of a noisy channel.

\subsection{A Simple Entanglement Assisted Code}

The simplest quantum error correcting code is the three qubit repetition code, which encodes a single qubit, and corrects against a single bit flip on any of the three qubits.  The qubit is encoded by using a Controlled-NOT gate (CNOT) on the state with each of the two ancilla qubits.  By adjusting the encoding procedure to use half of maximally entangled state, $|\Psi^+\rangle_{AB} = \frac{1}{\sqrt{2}}(|00\rangle + |11\rangle)$, instead of the pure state ancillas, and encoding the pure state, $|\phi\rangle_A = \alpha |0\rangle + \beta |1\rangle$, we obtain the codeword,
\begin{widetext}
\begin{eqnarray}
|\Phi \rangle \rangle &=& \frac{\alpha}{2} \Big( |000\rangle_A |00\rangle_B + |001\rangle_A |01\rangle_B + |010\rangle_A |10\rangle_B + |011\rangle_A |11\rangle_B \Big) \nonumber \\
&&+ \; \frac{\beta}{2} \Big( |111\rangle_A |00\rangle_B + |110\rangle_A |01\rangle_B + |101\rangle_A |10\rangle_B + |100\rangle_A |11\rangle_B \Big) \; ,
\end{eqnarray}
\end{widetext}
which can easily be seen to correct a single bit flip on the first three qubits of the codeword.  However, in addition, if any combination of the second and third qubits undergoes a phase flip, then these errors are also correctable.  For bit flip errors, we can see that the structure of the entanglement code depends on the labels attributable to the non-transmitted portion of the codewords, and we essentially break the coding subspaces down to a classical $(3,1,3)$ code for these labelled spaces.
If we look at the stabilizer formalism for the three qubit repetition code, we can see that the elements of the stabilizer act on the code space to take it to an orthogonal subspace by flipping the phases of the components of the logical zero and one.  For each of the codewords we see that the phases change as,
\begin{eqnarray}
111 &\rightarrow& ++++ \\
Z1Z &\rightarrow& +-+- \\
1ZZ &\rightarrow& +--+ \\
ZZ1 &\rightarrow& ++-- \; .
\end{eqnarray}
However, the code in the example is not a single error correcting, $t=1$, code as the single qubit error $Z11$ is a logical error on the codeword, and hence cannot be corrected.  The obvious candidate for a $k=1$ single error-correcting entangled code is the five qubit $k=1$ single error correcting code \cite{laflamme96}.  The codewords for this code can be generated using a pair of entangled ancillas using a local unitary operation on the encoded state and the local halves of the two entangled states.  The unitary transformation is determined by the change of basis,
\begin{eqnarray}
|000\rangle &\rightarrow& |000\rangle - |011\rangle + |101\rangle - |110\rangle \nonumber \\
|001\rangle &\rightarrow& |001\rangle + |010\rangle - |100\rangle - |111\rangle \nonumber \\
|010\rangle &\rightarrow& -|001\rangle + |010\rangle + |100\rangle - |111\rangle \nonumber \\
|011\rangle &\rightarrow& -\big(|000\rangle + |011\rangle + |101\rangle + |110\rangle \big) \nonumber \\
|100\rangle &\rightarrow& -\big(|001\rangle + |010\rangle + |100\rangle + |111\rangle \big) \nonumber \\
|101\rangle &\rightarrow& -|000\rangle + |011\rangle + |101\rangle - |110\rangle \nonumber \\
|110\rangle &\rightarrow& -|000\rangle - |011\rangle + |101\rangle + |110\rangle \nonumber \\
|111\rangle &\rightarrow& -|001\rangle + |010\rangle - |100\rangle + |111\rangle \; ,
\end{eqnarray}
which gives the codewords for the five qubit single error-correcting code.  The five qubit code is thus equivalent to an entanglement assisted three qubit code, with an unassisted stabilizer, $S = \{ 111, X1X, ZYY, YYZ\}$.  As the five qubit single error correcting code can correct single errors on the last two qubits of the codewords, and these qubits are noiseless in the entangled code, the code is not very efficient in maximizing the number of errors that can be corrected.  In order to examine the error correcting capability of linear entangled codes we must look again at the quantum Hamming bound.

\subsection{Revising the Quantum Hamming Bound}

The reason that entanglement assisted codes are better than their non-entangled counterparts, is that the entanglement allows us to increase the dimension of the decoding Hilbert space to $2^{2m+k}$ dimensions, for $m$ the number of entangled ancillas, compared to the $2^{m+k}$ dimensions for $m$ unentangled ancilla qubits.  This gives a revised quantum Hamming bound for entanglement assisted codewords as,
\begin{equation}
2^k \sum_{j=0}^{t} 3^j \left( \begin{array}{c} n \\ j \end{array} \right) \leq 2^{2n-k} \; ,
\label{eqn:entHamming}
\end{equation}
which is easily satisfied for the 3-qubit code above, with $k=1$, $n=3$, and $t=1$.

The asymptotic form of Eq.(\ref{eqn:entHamming}) is given by,
\begin{equation}
\frac{k}{n} \leq \frac{m}{n} + 1 - \frac{t}{n} \log_2 3 - H_2\left(\frac{t}{n}\right) \; ,
\end{equation}
where $m=n-k$, which can be seen to exceed the normal quantum Hamming bound.  Substituting the rate and error probability, for $R = k/n$, and $p = t/n$, we have,
\begin{equation}
R \leq 1 - \frac{p}{2} \log_2 3 - \frac{1}{2}H_2\left(p\right) \; ,
\end{equation}
which corresponds to the entanglement assisted quantum capacity for the depolarizing channel.

\subsection{General Entangled Linear Codes}

\begin{figure}
\epsfig{file=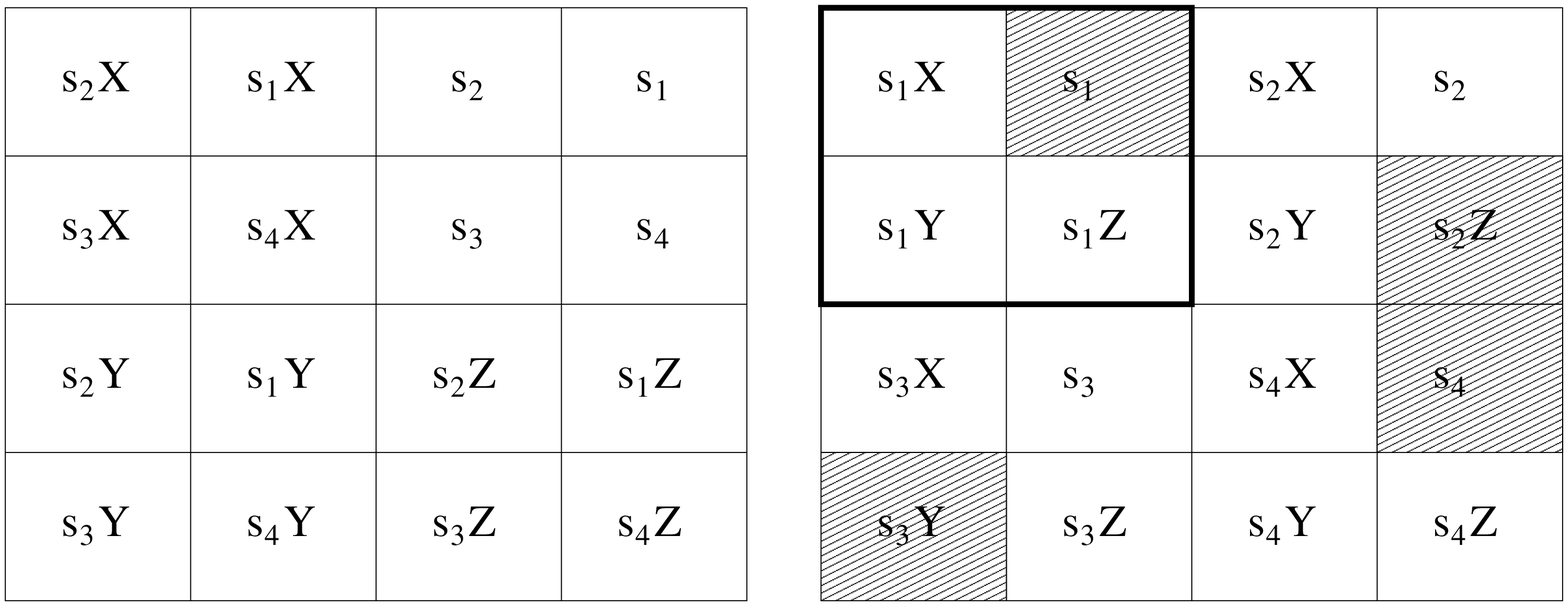,height=3.25cm}
\caption{The left hand figure represents the normalizer from FIG. \ref{fig:stab01}, subdivided into the individual elements of the stabilizer and its cosets.  Under an entangled code these all map to orthogonal subspaces, and hence the new stabilizer $S'= s_1$, and logical errors $X,Y$ and $Z$, form the normalizer of a new stabilizer.  The increased ability to choose correctable errors allows us to attain the entanglement assisted capacity for certain classes of qubit channels.\label{fig:stab02}}
\end{figure}

For general stabilizer codes we must prove that the elements of the stabilizer act to take the codewords to orthogonal subspaces, which are also orthogonal to the other error spaces generated by the elements outside the normalizer.
As a code is constructed by a unitary transformation, if we act on the states $|k\rangle \otimes |00...0\rangle$ and $|k\rangle \otimes |00...1\rangle$, for $|k\rangle$, $|l\rangle$, basis states for the state to be encoded, we find,
\begin{equation}
\langle k|\otimes \langle 0...00|U^{\dag}U|l\rangle \otimes |00...1\rangle = 0 \; ,
\end{equation}
and the code states with orthogonal ancillas are obviously orthogonal.  Hence, by introducing the set of bit flip operators, $P(X)$, and applying them to the ancillas, we then have,
\begin{equation}
\langle k|\otimes \langle 0...00|P_i(X)U^{\dag}UP_j(X)|l\rangle \otimes |00...0\rangle = \delta_{ij}\delta_{kl} \; ,
\end{equation}
which holds for all possible combinations of bit flip operators.  By combining these states with the basis, $\{ |k\rangle \}$, for the message qubits we then have a basis for the encoder's Hilbert space. When the ancillas consist of the shared entangled states, $|\Psi^+\rangle_{AB}$, then we encode a linear combination of orthogonal states,
\begin{equation}
|\tilde{k}\rangle = \sum_{P(X)} U_{A} \big( |k_{A}\rangle \otimes P(X)|00...0_{A}\rangle \big) \otimes P(X)|00...0_{B}\rangle ,
\end{equation}
where the sum is over the set of all possible bit flips on the ancillas.  Now, suppose, $E \in S$, is an element of the stabilizer, then,
\begin{equation}
\langle k|\otimes \langle 0...00|U^{\dag}EUP(X)|l\rangle \otimes |00...0\rangle = 0 \; ,
\end{equation}
for $P(X) \neq \mathbb{I}$, this is because $E=E^{\dag}$ for $E \in S$, and so acting to the left the error leaves the bra invariant.  This also applies for the encoding of all the basis states of the message, and so these states are orthogonal to the code space.  Also, the encoding operation has the freedom to ensure,
\begin{widetext}
\begin{eqnarray}
\langle \tilde{k}|E|\tilde{l}\rangle &=& \sum_{P_1(X),P_2(X)} \langle \mathbf{0}_B|P_2(X)\langle \mathbf{0}_A|P_2(X)\langle k_A| U^{\dag}_A E U_{A} |l_A\rangle P_1(X)|\mathbf{0}_{A}\rangle P_1(X)|\mathbf{0}_{B}\rangle \label{eqn:main1} \\
&=& \sum_{P(X)} \langle \mathbf{0}_A|P(X)\langle k_A| U^{\dag}_A E U_{A} |l_A\rangle P(X)|\mathbf{0}_{A}\rangle \label{eqn:main2} \\
&=& 0 \; ,
\label{eqn:main3}
\end{eqnarray}
\end{widetext}
where the final line follows from the fact that, since the states, $\{ P(X)|\mathbf{0}\rangle \}$, tensored with a basis for the space to be encoded, forms a basis for the total encoding space, we can choose the encoding, $U$, such that the encoded basis states form a basis consisting of the eigenvectors of $S$ (which is possible as all the elements of $S$ commute), ensuring half are $+1$ eigenvectors and half $-1$ eigenvectors of any, $E \in S$, excluding the identity.  For any two stabilizer elements, $E,F \in S$, the product, $E^{\dag}F = EF$, is also in the stabilizer, and hence the argument above shows that any two different elements of the stabilizer map codewords to orthogonal subspaces for the entangled code.

Furthermore, we can also choose the basis, such that for each given $P(X)$, the states $\{ U|k\rangle P(X)|\mathbf{0}\rangle \}$, all sit in the same eigenspace.  To see this, note that for the $2^{n-k}$ generators of the stabilizer, we write out binary strings with each element of the string corresponding to whether the given basis state is a $+1$ or $-1$ eigenstate of that element.  Each of these strings then corresponds to a given $P(X)$, and the remaining $2^k$ bits required to label the basis can then correspond to each of the $2^k$ basis states of the message space $|j\rangle$.  The fact that the subspaces generated by the stabilizer elements are then orthogonal to the subspaces generated by errors outside the normalizer can then be shown by substituting $EF = EFE^2$ into Eqs.(\ref{eqn:main1})-(\ref{eqn:main3}), and noting that the states for each term in Eq.(\ref{eqn:main2}) are both $\pm 1$ eigenvectors of $E$ with the same sign, and then noting that $EF = -FE$.  The proofs that there exist entangled stabilizer codes that attain the capacities for both unital qubit channels and the qubit erasure channel, with an amount of entanglement per channel given by, $\mathcal{E}_Q^{\mathrm{Random}} = 1 - Q_E$, are outlined in the Appendix.

\subsection{Entangled Codes and Degeneracy}

For an entangled linear code that encodes $k$ qubits using $m$ entangled ancillas and $a$ non-entangled ancillas, there are, $|\mathcal{C}| = 2^{n-k} = 2^{2m+a}$, copies of the code space, $|E| = 4^{k+m+a}$, physical errors, and, $|N(S)|/|S| = 4^k$, logical errors on each subspace, hence,
\begin{equation}
|S| = \frac{|E|}{|\mathcal{C}|\cdot 4^k} = 2^a \; ,
\label{eqn:size_of_S}
\end{equation}
and the number of elements in the stabilizer is determined by the number of non-entangled ancillas used for the encoding.  The case of every ancilla being an entangled state, $a=0$, reduces Eq.(\ref{eqn:size_of_S}) to, $|S| = 1$, and there are no degenerate errors for such an entangled code.  The parameter, $\mathcal{A}=a/n$, therefore gives a measure of the degeneracy possible with the encoding, as the size of the stabilizer scales as, $|S| = 2^{n\mathcal{A}}$.

\subsection{``Teleportation'' Codes}
\label{sec:tele_code}

The next entanglement assisted codes we look at are teleportation codes, based on the structure of the standard teleportation protocol.  The classical channel may be modeled by a ``classical'' quantum channel such as the total dephasing channel, $\Lambda\rho = \frac{1}{2}(\rho + \sigma^Z \rho \sigma^Z)$, which has classical capacities of, $C = C_E = 1$, but a zero quantum capacity.  The entanglement assisted quantum capacity of $\Lambda$ is therefore, $Q_E(\Lambda) = \frac{1}{2}C_E = \frac{1}{2}$.

First, we examine the standard teleportation protocol using this channel.  For a single entangled resource, and a single qubit, $|\psi \rangle = \alpha |0\rangle + \beta |1\rangle$, we undertake the transformation,
\begin{equation}
|00_{A}\rangle \otimes |\psi_{A}\rangle \otimes |\Psi^+_{AB}\rangle \rightarrow U_{A} |00_{A}\rangle \otimes |\psi_{A}\rangle \otimes |\Psi^+_{AB}\rangle ,
\label{eqn:tele_code_trans}
\end{equation}
where the unitary encoding operation is given by,
\begin{eqnarray}
U_{A} &=& \mathbb{I} \otimes \mathbb{I} \otimes |\Psi^+\rangle \langle \Psi^+| + \mathbb{I} \otimes \sigma_x \otimes |\Phi^+\rangle \langle \Phi^+| \\
&& + \; \sigma_x \otimes \mathbb{I}\otimes |\Phi^-\rangle \langle \Phi^-| + \sigma_x \otimes \sigma_x \otimes |\Psi^-\rangle \langle \Psi^-| . \nonumber 
\end{eqnarray}
The first two qubits in Eq.(\ref{eqn:tele_code_trans}) are sent through the quantum channel $\Lambda$, and the operation $V_{AB}$ applied to the three qubits at the receivers end of the channel, where,
\begin{eqnarray}
V_{AB} &=& |00_{A}\rangle \langle 00_{A}| \otimes \mathbb{I}_{B} + |01_{A}\rangle \langle 01_{A}| \otimes \sigma_{x B} \nonumber \\
&& + \; |10_{A}\rangle \langle 10_{A}| \otimes \sigma_{y B} + |11_{A}\rangle \langle 11_{A}| \otimes \sigma_{z B} \; ,
\end{eqnarray}
to give the resultant decoded state.  The codeword generated in Equation (\ref{eqn:tele_code_trans}) can be written explicitly as,
\begin{eqnarray}
|\Phi\rangle\rangle &=& |00\rangle |\Psi^+\rangle |\psi\rangle + |01\rangle |\Phi^+\rangle \sigma_x|\psi\rangle \nonumber \\
&&+ \; |10\rangle |\Phi^-\rangle \sigma_y |\psi\rangle + |11\rangle |\Psi^-\rangle \sigma_z |\psi\rangle \; ,
\end{eqnarray}
and we may note that the label states consisting of the Bell states that are not sent through the channel, can be considered redundant, and so we may encode the state by ignoring the ancillas, and using a CNOT followed by a Hadamard transformation on the encoded qubit.  Thus,
\begin{eqnarray}
|\Phi\rangle\rangle &=& |00\rangle |\psi\rangle + |01\rangle \sigma_x|\psi\rangle + |10\rangle \sigma_y |\psi\rangle \nonumber \\
&&+\; |11\rangle \sigma_z |\psi\rangle \; ,
\end{eqnarray}
and the label states $|00\rangle, |01\rangle, |10\rangle$, and $|11\rangle$, are each invariant, up to a global phase, under the action of the channel $\Lambda$.  Upon measurement of the first two qubits of $\Lambda\rho_{AB}$ we obtain an ``error syndrome'', which is then corrected by application of the appropriate unitary transformation.  The code thus requires two uses of the channel $\Lambda$ for each state sent, giving a rate $R = 1/2$, which attains the entanglement assisted capacity for this channel.  We can also note that the amount of entanglement required per channel is simply one e-bit per two channels or, $\mathcal{E}_Q = 1 - Q_E = 1/2$.  However, it is apparent that teleportation codes can only be optimal if, $C_E = C$, for the channel.  An example when they are not optimal is given a dephasing channel for, $p \neq 1/2$, where the teleportation code still only has a rate, $R = 1/2$, whilst entangled codes attain, $Q_E = 1-\frac{1}{2}H(p) > 1/2$.

There is, however, a notable difference between the teleportation code compared to the entangled linear codes for the $p=1/2$ dephasing channel, in that the channel created by the teleportation code is noiseless in the case of a finite number of uses of the channel, whereas the entangled linear codes obtain arbitrarily high fidelity only in the asymptotic limit.  In this sense the teleportation code is an analogue of a zero error code \cite{cover}, but this relies on the ability of the channel to transmit a zero error classical code.

\section{Entanglement per Channel Requirements}

The entanglement quantum Hamming bound for $m$ e-bits in a length $n$ code includes the terms, $n = k + m + a$, where $a$ is the number of unentangled ancillas required in the code.  Therefore,
\begin{equation}
2^k \sum_{j=0}^{t} 3^j \left( \begin{array}{c} n \\ j \end{array} \right) \leq 2^{2m+k+a} \; ,
\label{eqn:entHamming2}
\end{equation}
which in the asymptotic limit gives,
\begin{equation}
R \leq 1 + \mathcal{E} - p\log 3 - H(p) \; .
\end{equation}
Since, $m \leq n - k$, equality gives the entanglement assisted capacity for the depolarizing channel in Equation (\ref{eqn:entHamming}), and so this requires a minimum entanglement of, $\mathcal{E} = 1-R$, e-bits per channel to reach capacity with a non-degenerate code.  If the entanglement per channel is given by, $\mathcal{E} = \frac{1}{M}(1-R)$, then we obtain a family of entanglement Hamming bounds, where,
\begin{equation}
R \leq 1 - \frac{M}{M+1}\left( p \log 3 + H(p)\right) \; ,
\end{equation}
corresponding to the limits for non-degenerate codes with the given amount of entanglement per channel.
As the amount of entanglement per channel decreases the size of the stabilizer increases, and hence the possibility of using degeneracy in codes to increase the capacity beyond the corresponding Hamming bound is also presumed to increase.

Whenever $C_E > C$, the the entanglement assisted classical capacity is generally assumed to require, $\mathcal{E} = S(\rho)$, e-bits per channel in order to achieve the capacity, where $\rho$ is the state that achieves the maximum in Eq.(\ref{eqn:C_E}).  In the case of unital qubit channels, if standard dense coding is used with halves of shared maximally entangled pairs, which are then sent via entangled quantum codes, the capacity, $C_E = 2Q_E$, can be achieved with $\mathcal{E} = 1$.  This is not a very useful fact, as for this type of channel it is already well known that dense coding achieves this capacity, with $\mathcal{E}_{DC} = 1$, \cite{bennett96,bennett99}.  However, if a degenerate entangled code can be found with, $\mathcal{E} < 1 - Q_E$, then the capacity $C_E$ could be achieved with $\mathcal{E} < 1$.  With this in mind, we examine what bounds exist on the entanglement requirements $\mathcal{E}_Q$ and $\mathcal{E}_C$.

\subsection{Upper and Lower Bounds on the Required Entanglement}

Given $n\mathcal{E}_Q$ shared e-bits, we can simulate $nQ_E$ noiseless quantum channels with $n$ noisy channels.  Therefore, if we are given an extra $nQ_E$ shared e-bits, we can utilize dense coding with these extra e-bits to obtain a capacity of $nC' = 2nQ_E = nC_E$, and hence, $\mathcal{E}_Q \geq \mathcal{E}_C - Q_E$.  Obtaining a lower bound on $\mathcal{E}_C$ therefore gives a lower bound on $\mathcal{E}_Q$.  Similarly, given $\mathcal{E}_C$ e-bits per channel means we can simulate $nC_E$ noiseless classical channels with $n$ noisy quantum channels, hence with, $nC_E/2 = nQ_E$, e-bits of extra entanglement we can teleport to create $nQ_E$ noiseless quantum channels, and so we obtain the bound, $\mathcal{E}_Q \leq \mathcal{E}_C + Q_E$.  Hence, a lower bound on $\mathcal{E}_Q$ gives a lower bound on $\mathcal{E}_C$.  Putting this two inequalities together gives the relationship,
\begin{equation}
Q_E \geq |\mathcal{E}_C - \mathcal{E}_Q| \; ,
\label{eqn:bound1}
\end{equation}
which relates the capacities to the minimum entanglement requirements.  From this inequality it is easily seen that as the channel becomes so noisy that the entanglement assisted capacities become small, the entanglement requirements converge, that is, $Q_E \rightarrow 0 \Rightarrow \mathcal{E}_C \rightarrow \mathcal{E}_Q$.

When coupled with a noiseless channel, the capacity of any noisy quantum channel is additive \cite{bennett96}.  If we have $n$ copies of a noisy channel, and we add $m$ noiseless channels, such that, $m/n \simeq \mathcal{E}_Q$, then sending maximally entangled states through the noiseless channel will give us enough entanglement to achieve the entanglement-assisted capacity for the noisy channels.  Hence, $m+nQ \simeq n\mathcal{E}_Q + nQ \geq nQ_E$, and,
\begin{equation}
\mathcal{E}_Q \geq Q_E - Q \; ,
\label{eqn:bound2}
\end{equation}
giving a lower bound on the required entanglement in terms of the channel capacities.  Similarly, the classical capacity version of this inequality also applies, where, $\mathcal{E}_C \geq C_E - C$.
As the classical capacity of a channel is at least as great as the quantum capacity, the ratio, $Q/C \leq 1$.  Thus,
\begin{equation}
\mathcal{E}_C \geq \frac{Q}{C}(C_E - C) \; .
\label{eqn:lowerE_Cbound}
\end{equation}
The combination of Eqs.(\ref{eqn:bound2}) and (\ref{eqn:lowerE_Cbound}) give upper and lower bounds on the quantum capacity of a channel based on the entanglement capacities, classical capacity, and required entanglement, where,
\begin{equation}
\frac{\mathcal{E}_C}{C_E / C - 1} \geq Q \geq Q_E - \mathcal{E}_Q \; ,
\end{equation}
although the upper bound may not be very tight for many channels.

\subsection{Examples for Particular Channels}

The quantum erasure channel has known quantum and classical capacities \cite{bennett97,barnum98a}.  The entanglement assisted capacities are, $C_E = 2 - 2\epsilon$, and, $Q_E = 1 - \epsilon$, for a channel erasure probability $\epsilon$ \cite{bennett99}.
Using Eq.(\ref{eqn:bound2}) we can show that if, $\mathcal{E}_Q = \epsilon - \delta$, for $\delta > 0$, then this implies, $Q > 1 - 2\epsilon$, a contradiction.  Hence, we have a lower bound on, $\mathcal{E}_Q$, which combined with the random coding bound, $\mathcal{E}_Q \leq 1-Q_E = \epsilon$, gives the equality, $\mathcal{E}_Q = \epsilon$, for the erasure channel.  For the erasure channel we can therefore see that, whilst $Q_E$ is attainable with $\epsilon$ e-bits per channel, the classical capacity $C_E$ is only attainable with more than, $\mathcal{E}_C \geq 1-\epsilon$, e-bits (if not 1 e-bit), and so for $\mathcal{E} = \epsilon < 1/2$, e-bits per channel, the factor of two relationship between these capacities no longer holds, and hence, $C_{\mathcal{E}} < 2Q_{\mathcal{E}} = C_E$.

For the dephasing channel, $\Lambda = (1-p)\mathbb{I} + p Z$, for $Z$ a phase flip of the qubit, we can also calculate bounds on the required entanglement, where,
\begin{eqnarray}
1-H(p) \leq \mathcal{E}_C \leq 1 \\
\mathcal{E}_Q = \frac{1}{2}H(p) \; ,
\end{eqnarray}
where equality is obtained in the second case as the upper and lower bounds again coincide.

The entanglement assisted capacities for unital qubit channels are determined by sending half of the maximally entangled state through the channel \cite{bennett96,bowen01b}.  This gives a lower bound on the entanglement of the Bell diagonal state generated by sending half the maximally entangled state through the channel, where,
\begin{equation}
E(\rho) \geq C_E - 1 = 1 - S(\rho) \; ,
\end{equation}
and this quantity is equivalent to the distillable entanglement of $\rho$ using the Hashing protocol.  This bound is derived for, $\mathcal{E}_Q = 1 - Q_E$, however if, $\mathcal{E}_Q = 1 - Q_E - \delta$, for $\delta > 0$, then the lower bound on the entanglement is higher.  This is because obtaining the capacity with less entanglement per channel requires degeneracy in the code, and the degeneracy is what allows us to beat the Hashing bound \cite{divincenzo98}.

Finally, note that for entanglement breaking channels the two entanglement assisted capacities are bounded by $C_E \leq 1$ and $Q_E \leq 1/2$, otherwise the ratio of entanglement that can be sent through the channel $E' = Q_E$ and the initial entanglement $E = 1-Q_E$ would be larger than one, allowing us to create entanglement through the channel.

\section{Discussion}

At this point we make the conjecture that the unassisted quantum capacity of a channel is given by,
\begin{equation}
Q = Q_E - \mathcal{E}_Q = C_E - \mathcal{E}_C \; ,
\label{eqn:conjecture}
\end{equation}
whenever, $Q_E \geq \mathcal{E}_Q$, (and both, $C_E \geq \mathcal{E}_C$, and, $C_E > C$, in the second equality), and zero otherwise.  The first equality holds for the dephasing and erasure channels, and the second equality will hold for both the dephasing and erasure channels provided, $\mathcal{E}_C = 1$, and, $p\neq 1/2$.  The second equality requires the condition $C_E > C$, and would imply the equality, $Q_E = \mathcal{E}_C - \mathcal{E}_Q$.  If this conjecture is true, then calculating the capacity of a quantum channel could be achieved by calculating $C_E$ from Eq.(\ref{eqn:C_E}), and either $\mathcal{E}_C$ or $\mathcal{E}_Q$.  Calculating either of these two quantities, however, may be just as difficult as determining the unassisted quantum capacity itself.  So far it has been shown in this paper that,
\begin{eqnarray}
Q &\geq& Q_E - \mathcal{E}_Q \\
C_E - \mathcal{E}_C &\geq& Q_E - \mathcal{E}_Q \; ,
\end{eqnarray}
so the reverse inequalities need to be shown for both of these equations to prove the conjecture.  The second of these is likely to be a problem, as it breaks down whenever $C_E = C$, as this implies $Q=C$, which is not true for many known channels.  However, we do have the relationship,
\begin{equation}
Q \geq \frac{Q}{C}C_E - \mathcal{E}_C \; ,
\end{equation}
for $Q > 0$, that gives, $Q \geq Q-\mathcal{E}_C$, whenever $C_E = C$.

The existence of degenerate entangled codes for unital qubit channels would also imply that the entanglement-assisted classical capacity for such channels could be achieved with an amount of entanglement per channel, $\mathcal{E}_C < 1$.  This would be a surprising result, as it is well known that dense coding achieves the capacity with, $\mathcal{E} = 1$, and this protocol has been assumed to be optimal.

\section{Conclusion}

In this paper bounds on the minimum amount of shared entanglement necessary per channel required to achieve the entanglement assisted capacities were derived.  An upper bound on the entanglement required, for classes of qubit channels, was obtained by introducing entanglement-assisted linear quantum codes.  The difference between the amounts of entanglement required were shown to vanish as the entanglement assisted capacities became small.

It was then shown that the unassisted capacities of the channel were bounded from below by the difference in the entanglement assisted capacities and the amount of entanglement required to achieve them.  The introduction of degeneracy into these entanglement assisted codes would therefore give a lower bound on the unassisted capacity for some of these channels that is higher than currently known lower bounds.  The use of such codes would also allow the entanglement assisted classical capacity, for classes of unital qubit channels, to be attained with less than one e-bit per channel.  Whether or not the generation of degenerate entangled codes will be easier than simply determining classes of unassisted degenerate quantum codes is not known, but this method does provide a second avenue for investigation.

Finally, a conjecture was made that there exists an equality between the unassisted quantum capacity of a channel and the difference in the entanglement assisted capacity and the respective minimum required entanglement attaining that capacity.  If this conjecture is shown to hold, then it provides a further link between entanglement as a resource and quantum communication.

\appendix*
\section{Summary of the proof for entanglement assisted codes achieving $Q_E$}

By taking random stabilizer codes, and showing that the average probability of error can be made vanishingly small in the limit of large block sizes, we may infer the existence of stabilizer codes that have a vanishingly small maximal failure probability, with rates arbitrarily close to the capacity.  Firstly, we outline the case of the unital qubit channels.
For such a channel, the number of typical errors with total probability bounded by, $P \geq 1-\eta$, for $\eta >0$, is bounded above by,
\begin{equation}
N \leq 2^{nS\big((\mathbb{I}\otimes\Lambda) |\Psi^+\rangle\langle\Psi^+|\big)+2n\delta}\; ,
\end{equation}
for any $\delta > 0$, with $n$ sufficiently large.
The total probability of failure is then given by,
\begin{eqnarray}
P(\mathrm{Fail}) &\leq& 2^{nS\big( (\mathbb{I}\otimes\Lambda) |\Psi^+\rangle\langle\Psi^+|\big)+2n\delta}2^{-(2n-2k)} + \eta \nonumber \\
&\leq& 2^{n\big[S\big( (\mathbb{I}\otimes\Lambda) |\Psi^+\rangle\langle\Psi^+|\big)+2R-2+2\delta\big]} + \eta ,
\end{eqnarray}
where the first term gives the probability of two typical errors having the same syndrome, and the second term is the probability of an atypical error.  The number of syndromes is determined by the fact that there are $2^{2n-k}$ dimensions divided amongst $2^k$ encoded qubits.
Hence, the average probability of failure becomes arbitrarily small, for large $n$, for any, $R < 1 - \frac{1}{2} S(\rho)-\delta$, where $\rho = (\mathbb{I}\otimes\Lambda) |\Psi^+\rangle\langle\Psi^+|$.  As the rate for the average over stabilizer codes attains rate $R$ there must exist particular codes with rate $R$ and arbitrarily small average failure probability.  Expurgation, that is removal of half of the codewords corresponding to the highest probabilities of error, may then be used on such codes to assure that the maximal probability of block error for such a code is also arbitrarily small, with minimal effect on the rate of the new code \cite{mackay}.  Thus the capacity is given by,
\begin{equation}
Q_E = 1 - \frac{1}{2} S(\rho) \; ,
\end{equation}
for entanglement-assisted codes using, $\mathcal{E} = \frac{1}{2}S(\rho)$, e-bits per channel.

In the case of the qubit erasure channel, with erasure probability $\epsilon$, the location of each of the errors is known.  For large $n$ the number of typical errors approaches, $4^{n\epsilon}$, giving a average failure probability,
\begin{eqnarray}
P(\mathrm{Fail}) &\leq& 2^{2n\epsilon+2n\delta}2^{-(2n-2k)} + \eta \nonumber \\
&\leq& 2^{2n(\epsilon-1+R+\delta)} + \eta \; ,
\end{eqnarray}
which vanishes for large $n$, provided, $R < 1-\epsilon-\delta$, for any, $\delta >0$.  Hence, the capacity obtained in this case is,
\begin{equation}
Q_E = 1 - \epsilon \; ,
\end{equation}
using, $\mathcal{E} = \epsilon$, e-bits per channel.

The amount of entanglement required in both these cases stems from the number of entangled states, $m = n - k$, utilized in the code.  Dividing through by the number of channels $n$ we have an amount of entanglement, $\mathcal{E} = 1 - R > 1 - Q_E + \delta$, per channel in the random coding derivation.  Hence, in these cases the required entanglement is given by, $\mathcal{E}^{\mathrm{Random}}_Q = 1 - Q_E$.

\begin{acknowledgments}
Thanks to Luke Rallan and Sougato Bose for very helpful discussions.  GB is supported by the Oxford-Australia Trust, the Harmsworth Trust, and the CVCP.
\end{acknowledgments}


\end{document}